\documentclass{aa}
\usepackage{psfig}

\usepackage{graphicx}
\usepackage{sidecap}

\begin{document}
   \title{Indirect Imaging of an Accretion Disk Rim in the 
Long--period Interacting Binary W Crucis\thanks{Table 1
is only available in electronic form at the CDS via anonymous
ftp to \texttt{cdsarc.u-strasbg.fr} (130.79.128.5) or via
{http://cdsweb.u-strasbg.fr/Abstract.html}}}

   \author{K. Pavlovski
          \inst{1}
          \and
          G. Burki\inst{2}
          \and
          P. Mimica\inst{3}}

   \offprints{K. Pavlovski: pavlovski@sirius.phy.hr}

   \institute{Department of Physics, University
              of Zagreb, Bijeni\v{c}ka 32, 10$\,$000 Zagreb, Croatia
         \and
             Observatoire de Gen\`{e}ve, 51 ch.\ des Maillettes, 
             1290 Sauverny, Switzerland
          \and
          Max-Planck-Institut f\"{u}r Astrophysik, Postfach 1312, 
          85741 Garching, Germany }

\date{Received }

\abstract
 {Light curves of the long-period Algols are known for their
  complex shape (asymmetry in the eclipse, light variations outside
  eclipse, changes from cycle-to-cycle), and  their interpretation is
  not possible in the standard model of binary stars.}
 {  Complex
  structures present in these active Algol systems could be studied
  with the eclipse-mapping method which was successfully applied to the
  new 7-color photometric observations in the Geneva system of W Cru,
  belonging to the isolated group of these active Algols.}
 { Several cycles of this long-period (198.5 days) eclipsing binary
  have been covered by observations. We have used a modified Rutten's
  approach to the eclipse-mapping. The optimization of the system's
  parameters and the recovery of the disk intensity distrubution are
  performed using a genetic algorithm (GA).}  
  {It is found that a
  primary (hot) component is hidden in the thick accretion disk which
  confirms previous findings. The mass of the primary component,
  $M_{1} = 8.2$ M$_{\odot}$ indicates that it is a mid-B type star.
  The mass-losing component is filling its critical lobe which, for
  the system's parameters, means it is a G-type supergiant with a mass
  $M_{2} = 1.6$ M$_{\odot}$.  The disk is geometrically very extended
  and its outer radius is about 80\% of the primary's critical lobe. A
  reconstructed image reveals a rather clumpy and nonuniform
  brightness distribution of an accretion disk rim in this almost
  edge-on seen system. This clumpiness accounts for light curve
  distortions and asymmetries, as well as for secular changes.}
  {}


\authorrunning{K. Pavlovski, G. Burki \& P. Mimica}

\titlerunning{Indirect Imaging of W Cru}

\maketitle 

\section{Introduction}

W Cru (HD~105998) is a member of the rather sparse group of strongly
interacting binaries which are believed to be either in the first and
rapid phase of the mass exchange between components, or have just
recently gone it. While mass reversal has already happened there are
still many signatures of the almost cataclysmic events that were
taking place. Studying the UV spectra of some long-period binaries
taken with the spectrograph on-board IUE satellite Plavec \& Koch
(1978) have found numerous  emission lines of highly-ionized 
species. It was
immediately obvious that the underlying stellar photospheres of these
binaries are not hot enough to excite such features, and therefore the
accretion processes were invoked as an explanation. 
This sparse group of the long-period interacting binaries comprises
 SX Cas, RX Cas, W Cru, W Ser, V367 Cyg, and $\beta$ Lyr, and a few others 
less studied systems. 
 All these
binaries are well known for their active nature with large period
changes, almost permanent Balmer line emission features, peculiar
light curves, etc.\ (Plavec 1980).

\begin{figure}[]
\sidecaption
\psfig{file=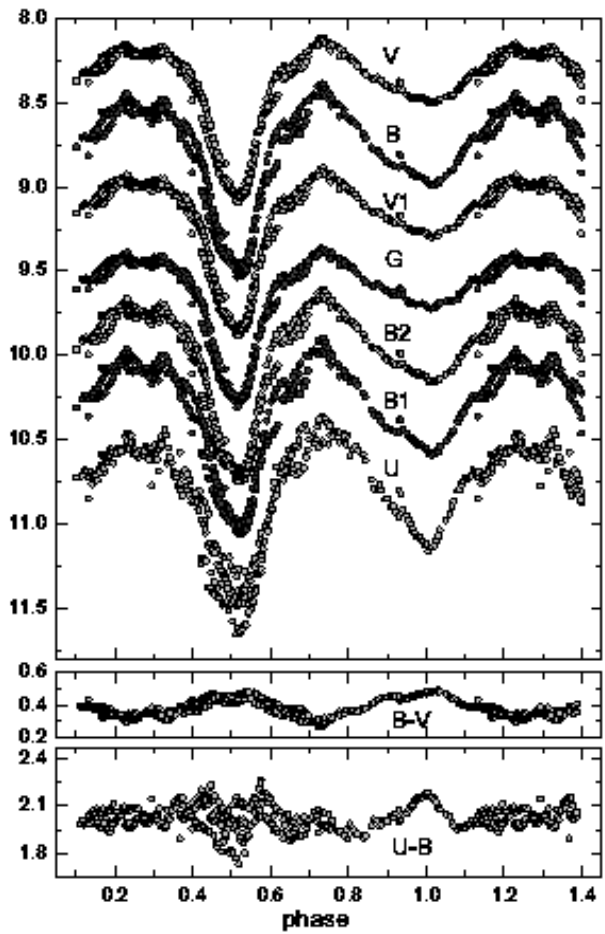,width=9.5cm}     
\caption{The light and colour curves  of W Cru in the Geneva 
photometric system. The light curves in $G$ and $B1$ have been
vertically shofted by 0.3 mag for clarity.}
\end{figure}

In the long-period Algol systems with $P > 6$ days, the primary
component is rather small relative to the binary separation and its
Roche lobe, and mass transfer results in a classical accretion disk.
Typically, this disk is a permanent structure characterized by strong
double-peaked emission lines in the spectrum. Since the disk hides the
mass-gaining component, which in these post mass-exchange binaries is
more massive and hotter than its companion, we are encountering a
paradox that the component expected to be more luminous 
 is invisible. This effect considerably affect the light curves.

 W Cru is not an unique binary system which harbors such a large
accretion disk. Several other long-period semi-detached binary 
systems have been found for which light curve solution has been 
found only after an optically and geometrically thick disk has 
been taken into account. Good examples are, RZ Oph (Zo{\l}a (1991),
and UU Cnc (Zo{\l}a et al.\ 1994), with the periods of 262 and 96
days, respectively.     
In all these long-period Algols the separation between the components
 is large, and hence, mass-losing components are supergiants, and disks
 which surround mass-accreting (hot and more massive) component 
typically extend several hundred solar radii in radial direction.

It is well known that light curves of the 'active algols' cannot be solved
with standard models that deal only with two (stellar) components. In this
respect the puzzling case of $\beta$ Lyrae is well-known (c.f.~Linnell 2000,
Harmanec 2002, and references therein) but the situation with other
binaries of this class is quite similar. A real break-through came with
Huang's (1963) disk--model for the $\beta$ Lyr, and its first application
to V356 Sgr by Wilson \& Caldwell (1978). Following this development Pavlovski
\& K\v{r}\'{i}\v{z} (1985) made a computer code for the synthesis of
light curves of close binary stars in which the influence of an accretion
disk is taken into account. They applied it in an analysis of the light
curves of SX Cas. Derived disk properties were in general accordance with
multi-wavelength studies by Plavec et al.\ (1982). Therefore, subsequent
studies have been performed on more difficult cases: RX Cas (Andersen,
Pavlovski \& Piirola 1989) and V367 Cyg (Pavlovski, Schneider \& Akan 1992).

In the meantime, some other attempts along the same disk-model but
with more or less physics details and various optimization techniques
have been undertaken. In particular, Zo{\l}a (1991) has implemented a
disk-model in the widely used Wilson-Devinney program, while Daems
(1998) introduced an analytical disk model.  However, these studies do
not take into account the fact that the accretion disk intensity
distribution is generally asymmetric. This means that a more realistic
model(s) is needed which is capable to account for erratic changes in
the light curves, asymmetry in ascending and descending branches,
unequal brightness of maxima, etc. In other words, a different
strategy is needed.

In the present paper we apply the technique of `eclipse mapping' (Horne 1985)
for the study of the complex light curves of the active Algol W Cru. The binary
system and new 7-color photometric observations in the Geneva system will be
described in Sec.\ 2 and 3, respectively. In Sect.\ 4 we describe the
model and calculations, while Sect.\ 5 is devoted to the discussion
of our results.

\section{Description of W Cru}

The binary star W Cru (HD~105998, CoD -58\degr4431) has the longest
period among the group of W Serpentis binaries. Its exceptionally long
period, $P = 198$ d, makes its a difficult object to observe. Only a
single spectroscopic study is published (Woolf 1962), and long
systematic photometric observations are scarce. However, the situation
was considerably improved after Plavec's (1984) appeal for new
photometric observations. As a result two photoelectric data sets of W
Cru were secured at the Auckland Observatory in 1984 and 1985 (Marino
et al.\ 1988), and at the Nigel Observatory from 1985 to 1991 (Pazzi
1993). In 1985 a systematic photoelectric campaign of W Cru was
initiated  by Z.\ Kviz at the Swiss Telescope on La Silla. As an
initial results, using Geneva photometry, a new determination of the
primary minimum was communicated by Kviz \& Rufener (1988). This aided
proper phasing of the satellite observations.

Woolf (1962) has observed W Cru spectroscopically throughout the
orbital cycle and detected only the presence of a less massive component
with supergiant characteristics of the spectral type G1Iab. Its RV
curve is well defined and enable Woolf to derive the system's mass
function $f(m) = 5.83$ M$_\odot$. Superimposed on the supergiant
spectrum are hydrogen Balmer emission lines and displaced absorption
lines; very typical for the stars of this class (c.f. Struve 1944).
Woolf recognized that the deeper minimum in the light curve is due to
an eclipse of the supergiant by an opaque body. Apparently, a more
massive star in this system is `invisible' - Woolf drew similarity of
W Cru to its more famous counterpart $\beta$ Lyrae.

While known for the other members of the group of these long-period
interacting binaries, period change is not yet found for W Cru
(c.f.~Zo{\l}a 1996). Asymmetry in the primary minimum, and
 somewhat erratic light curve from the cycle-to-cycle as is
seen in our Fig.~3, are, at least, some of the reasons for 
unconclusive results regarding this issue.

An important step in the understanding of this system was made by the
independent studies of Zo{\l}a (1996) and Daems (1998). Zo{\l}a (1996)
concentrated on the analysis of the light curves secured by Marino et
al. (1988) and Pazzi (1993), respectively, while the study by Daems
(1998) was more elaborated, and besides an independent modeling of
unpublished Geneva 7-color photometry, also included some new
spectrophotometric and spectroscopic observations. In particular,
Daems was able to construct the spectral energy distribution of W Cru
from IUE UV region up to IRAS FIR wavelengths. In both studies
disk-model was employed, Zo{\l}a used an cylindrical $\alpha$--disk
model, while Daems had applied an analytical torus-like disk. Further
discussion of their results is left for Sect.\ 5 where they will be
compared to the results of the present work.

\section{Photometric data and variability of the components}

 Geneva 7-colour photometric measurements (Golay, 1980; Rufener\& Nicolet, 
1988)
of W~Cru were obtained from December 1984 to  July 1989, using the Swiss
70~cm telescope at La Silla Observatory (ESO, Chile), equipped with the
photoelectric photometer P7 (Burnet \& Rufener, 1979). During this period,
378 measurements of weight $q \geq 1$ have been obtained (see Rufener, 
1988,
for the definition of the weight q). The photometric reduction procedure
was described by Rufener (1964, 1985); the photometric data in the Geneva system
are collected in the General Catalogue (Rufener, 1988) and its up-to-date
database (Burki, 2006). In the case of W~Cru, the all-sky reduction was
improved by the measurement of a neighbouring comparison star, HD~101021 
(K1III),
a well measured standard, with a confirmed stability: 440 measurements over
a period of 20 years, mean $V$ magnitude 5.1374 with a standard deviation
$\sigma_{V}$ = 0.0046. The same value of $\sigma_{V}$ can be adopted for the
global precision of the measures on W~Cru. The data on W~Cru are listed in 
Table~1
available at the CDS.

The magnitudes in each of the seven filters are obtained from the visual
magnitude $V$ and the six colour indices in the following manner :
\begin{equation}
i = V - [V - B] + [i - B]
\end{equation}
with i representing one of the seven filters $U$, $B$, $V$, $B_{1}$, $B_{2}$,
$V_{1}$, G. Remember that the Geneva $[U - B]$ and $[B - V]$ indices are not
normalized to zero for an A0V star as it is the case for the Johnson UBV
indices.

In Fig.~1 light curves in all seven photometric pass bands of the
Geneva system are shown. The following ephemeris is used for
the calculations of the phases: Min I (HJD)$ = 2440731.84 + 
198.537 \times E$  after Kohoutek (1988) and the
elements are derived from all available minima found in the
literature. 
The primary  minimum, to which phase $\varphi = 0.50$ is assigned 
throughout in this paper, and which is caused by
the eclipse of the visible component, became deeper for the
shorter wavelengths. However, light curve in the $U$-band has much larger
intrinsic scatter - large changes in the width and depth of the
primary minimum are evident for the light curve in the $U$-band than
for the other pass bands.  Unfortunately, phase coverage of the
secondary minimum is much poorer than primary and cycle-to-cycle
variations in the secondary minimum are not evident as for the primary
minimum. In fact, proper phase coverage in the secondary minimum has
been secured in a single cycle. The width of the secondary minimum
when an opaque disk is eclipsing its stellar companion is larger than
the width of the primary minimum. At shorter wavelengths the
secondary minimum becomes not only deeper but evidently narrower.
Moreover, some other peculiarities in the light curves of W Cru are
clearly seen: maximum following deeper minimum is higher than
preceding, there are asymmetries in descending and ascending branches
of both, primary and secondary, minima. Bumps and/or humps in the
light curves at the phases $\varphi = 0.2, 0.3, 0.6$ are very
pronounced, and, moreover, variable from cycle-to-cycle.

The color variations (bottom pannels in Fig.~1) are indicative 
of temperature changes.
In quadratures when the supergiant is seen from the side $[B-V]$ is
the largest. The amplitude of the changes in [B-V] index is about 0.25
mag. As is evident from Fig.~1 changes in $[U-B]$ are twice as large
then in $[B-V]$, i.e. about 0.50 mag. The variations are more
difficult to interpret in $[U-B]$.  Large and erratic changes in
$[U-B]$ around the primary minimum are present as consequence of a large
activity in this system on time scale comparable to system's orbital
period.  It seems that the source of these rapid fluctuations, which
has different amplitudes from cycle-to-cycle is screened in the phase
of the secondary minimum. But due to lack of more intensive phase
coverage our conclusion is rather tentative at this point (see also
Fig.~3).

\section{Eclipse-mapping: model and calculation}

\subsection{An outline of the model}

The eclipse-mapping technique introduced by Horne (1985) in
combination with Doppler tomography (Marsh \& Horne 1988) became an
essential tool for the study of the structure and mass-flow processes
in cataclysmic variables (CVs). In particular, eclipse--mapping is
used to reconstruct the surface brightness distribution of the
accretion disks in CVs, and is often assumed to be an indirect imaging
method.  Initially, only the information contained in the shape of the
eclipse was used in the assembly of the map of the accretion disk
surface brightness distribution.  However, other sources of the
radiation have been completely ignored, e.g. the secondary (often
cool) component. Rutten (1998) has improved the technique with the
development of 3D eclipse--mapping which fits the whole light-curve
instead of just the eclipsed part.

In the present study we have followed Rutten's (1998) approach. But,
unlike the case of CVs in Algols here accretion disk is not prevailing
source of radiation. Moreover, the hot component has a relatively
large radius and its screening of the accretion disk should be taken
into account.

Previous studies of the light curves of W Serpentis binaries have
shown that hot components might be exposed to various degrees,
Therefore, in the construction of the binary model we have started
from the general case in which all three main radiative sources are
presented: hot mass-accreting component surrounded by an optically
thick disk and cool mass-loosing component which (canonically) is
filling its Roche lobe.

The stellar surfaces are approximated by equipotential surfaces. These
surfaces are determined by two parameters: the mass ratio $q$ and the
filling factor $S$ (for description of the gravitational potential of
two rotating stars in binary system see e.g.  Hilditch 2001). We
approximated an equipotential surface with a grid of triangles. This
was done to avoid ray--tracing when calculating visibility. Geometry
of an accretion disk in our model is given by an $\alpha$--disk model,
i.e.~the disk is determined by the disk radius $R_{\rm d}$ and an
opening angle $\beta$.

The temperature distribution on stellar surfaces is parametrized by
the effective temperatures of the components. This constant
temperature distribution is modified by two effects: the gravity
darkening and the reflection effect.  For the stellar atmosphere in
radiative equilibrium the gravity darkening exponent is 0.25 (von
Zeipel 1924) while for the convective atmosphere is set to the value
0.08 according to Lucy (1968). Since we are using triangles rather
than a grid of points it is assumed that the acceleration due to
gravity on the stellar surface can be considered constant along the
triangle's surface for a sufficiently small triangle, and is equal to
the acceleration in one of its vertices. Temperature distribution on a
disk surface is computed in a simple power--law fashion. This
distribution is independent of the vertical coordinate, i.e.~there is
only radial distribution of the disk temperature.  For $\alpha$-disk
model the temperature of disk is $T \infty R^{-3/4}$.

 Although the geometry of an accretion disk is axially symmetric,
its temperature distribution, in general, is not. In the first phase
of the minimization (see Sec.\ 4.3) it is axially symmetric and
is a power-law in radius. However, in the second phase of the
minimization the requirement of the axisymmetry and radial dependence
is relaxed and the temperature distribution is allowed to be
essentially asymmetric. In all cases, the distribution is independent
of the vertical coordinate, i.e.~there is only radial and angular
(with respect to the disk axis) distribution of the disk temperature.
For $\alpha$-disk model the temperature of disk is $T \infty
R^{-3/4}$.

Radiation from one component influences the temperature distribution
on the other component. To increase the speed of the calculations the
reflection effect is computed in an approximate way. Stars are assumed
to be spheres of constant effective temperature. For example, when
calculating the effect of reflection of radiation from the secondary
(component 2) on the primary (component 1), a triangle whose one
vertex has the position $P(x,y,z)$ is considered. The radiation
dilution factor is defined by $ W = A\Omega /4\pi = A\,A_2 / 4\pi
d^2$, where $\Omega$ is the solid angle within which the inner
hemisphere of the secondary component is seen. The hemisphere has an
area $A_2$, and $d$ is the distance between the triangle and the
center of the secondary component. $A$ is the albedo of the primary
component, $A=0$ means total absorption of radiation, $A=1$ means
total reflection.

\subsection{Light curve synthesis}

The light curve is synthesized using a two-dimensional pixel grid onto
which the image of binary system at a particular position is
projected. This subsection describes how this is done in our code.

One of the biggest difficulties in computing the total flux from the
binary system at a given orbital phase and inclination is the problem
of determining which parts of the system are visible and should be
taken into flux determination.  One approach is to approximate each
surface element by a single point, then trace the ray of light from
that point to observer and check whether it intersects any other
surface on its way. While this method is as accurate as the
point--approximation, it is relatively slow since it involves
three--dimensional lines.

We have taken a different approach. The approximation of stars and
disk with a network of triangles is ideal when projecting a binary
system image onto a two-dimensional pixel grid since a triangle can be
filled using a simple algorithm (Bresenham 1965) which does not
involve neither multiplication nor division.  Triangles that are
facing away from observer are not considered. The remaining triangles
are sorted by distance and projected in descending order onto pixel
grid. If the closer triangle partially covers the further triangle, it
simply `overwrites' the other triangle's pixels with its own pixels.
What remains after all triangles have been projected is a
two--dimensional image of the binary system. Thus, the visibility
problem is solved.

\begin{figure}[]

\psfig{file=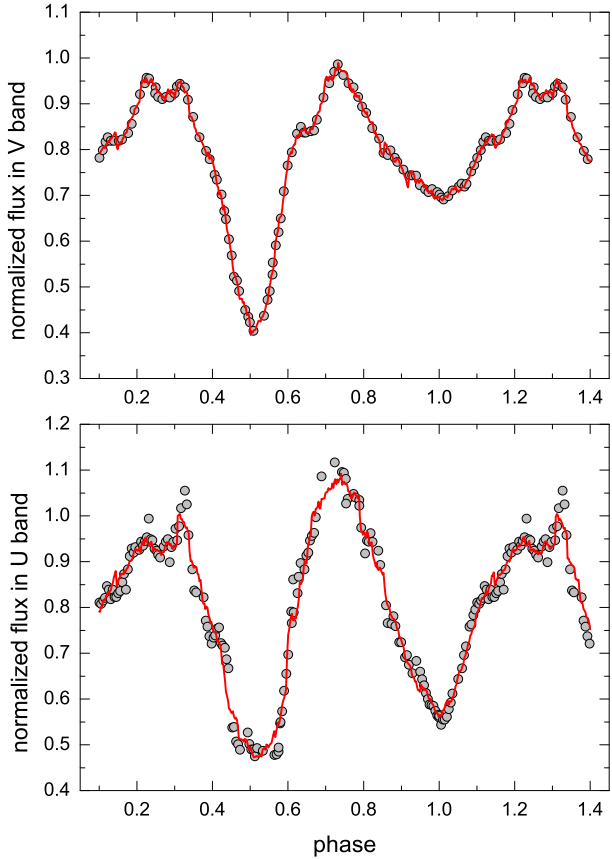,width=9.5cm}
\caption{In the upper panel the synthetic light curve  (solid line) calculated
for the optimized disk image and system's parameters compared to the
observed light curve of W Cru (circles) in the $V$-band of Geneva
system. In the bottom panel is the same but for the $U$-band of
the Geneva system. The light curves cover cycle 33 counting after the
epoch in Kohoutek's ephemeris (see Sec.\ 3).}
\end{figure}

Since only normalized light curves are used in our code, we define the
flux from a single triangle as the product of its surface and the
blackbody intensity for the triangle temperature. At a given orbital
phase the total flux from all the visible pixels is divided by the
total flux at phase 0.25.

We also include the possibility of having a source of this ``third
light,'' i.e.\ an unobscured component which provides a constant
contribution to the light curve in all orbital phases.

\subsection{Minimization by Genetic algorithm}

Each synthetic model of the binary system with an accretion disk can
be characterized by its light curve which is observed by an observer
located at a large distance from the binary system's center. In order
to deduce the parameters of the binary as well as the details of the
accretion disk temperature distribution one needs to find that
synthetic model whose light curve best matches the observed light
curve. In practice this means using a minimization procedure on a set
of parameters which describe the model.  In this work we have employed
the genetic algorithm (GA), see e.g.~Charbonneau (1995), as the
minimization procedure. GA is very efficient in case of a large number
of parameters and especially in case of the \emph{image
  reconstruction}.

The basic idea is to consider a set of $N$ binary system models which
are described by their basic stellar (mass ratio, Roche lobe filling
factors, stellar effective temperatures, inclination) and accretion
disk parameters (radius, height) as well as the accretion disk
temperature distribution (image). From the standpoint of a general
GA, each model is simply a set of real numbers which we call
\emph{organism} in the rest of this section.

In general, we define the \emph{fitness} $\chi_i$ of an organism as a
continuous monotonous function of its parameters. In our particular
case we define the fitness as the function of the light curve
synthesized from the organisms parameters and the observed light
curve. Since it is desirable that the fitness increases as the
difference between the two light curves decreases, we define fitness
as
\begin{equation}
\chi_i^2=\left[\sum_{j=1}^M (s_i(j)-s_{\rm obs}(j))^2\right]^{-1}\, ,
\end{equation}
where $s_{\rm obs}$ is the observed light curve containing $M$ points
and $s_i$ is the light curve synthesized from the organism $i$ at
exactly the same orbital phases present in $s_{\rm obs}$.

The GA starts with the random population of $N$ organisms which are
sorted according to their fitness. The most-fit (better) half of the
population is selected for reproduction, so that $N/4$ ``parent''
pairs are randomly selected (with better organisms being more probable
to be selected than worse (less-fit) organisms) and then two
``children'' from each pair are created to replace the worse half of
the population. This ``reproduction'' process is performed analogously
to the similar process in biology, i.e.~the organisms are divided into
chromosomes and then the chromosomes of the children are obtained from
the chromosomes of the parents through \emph{crossover} process: first
child gets one part of the chromosome from the first parent and the
second part from the second parent, while the second child gets the
first part from the second and the second part from the first parent.

This process converges very quickly to the nearest local minimum. In
order to avoid the ``stagnation'' of the population in a local minimum
random ``mutations'' are introduced, such as ``copy-errors'' during
crossover, random changes of the individual genes (numbers), as well
as the global mutations such as the ``asteroid hit'': a large number
of the population is radnomly re-initialized.

Our fitting process consists of two steps:
\begin{enumerate}
\item GA minimization of the binary system and accretion disk
  parameters without minimizing the accretion disk image (which is
  assumed to be a power-law function of radius);
  
\item GA minimization of an accretion disk image while keeping the
  binary system's parameters constant.
\end{enumerate}

This splitting of our method is necessary due to the limited
computational resources. A more general approach would involve an
intrinsically three-dimensional disk geometry and would then
simultaneously minimize both the geometry and the temperature
distribution. However, the number of parameters that describe disk
geometry would be much larger than in our approach. Unfortunately,
these parameters would only make a difference in a relatively small
part of the light curve near the minima, i.e. the amount of deviation
from the symmetric shape. Therefore, one would need to use a
prohibitively large number of organisms with a much larger set of
parameters to minimize the light curve. Instead, we use our two-phase
approach where the first phase can be thought of as the zeroth order
phase (fitting of the light curve with axisymmetric disk temperature
distribution and an optional hot spot), whereas the second phase can
be thought of as the first order correction (asymmetries in the disk
temperature distribution introduce irregularities in the symmetric
light curve minima). We find this approach more practical at this
time, due to the abovementioned limited computational resources.

\begin{figure*}
\psfig{file=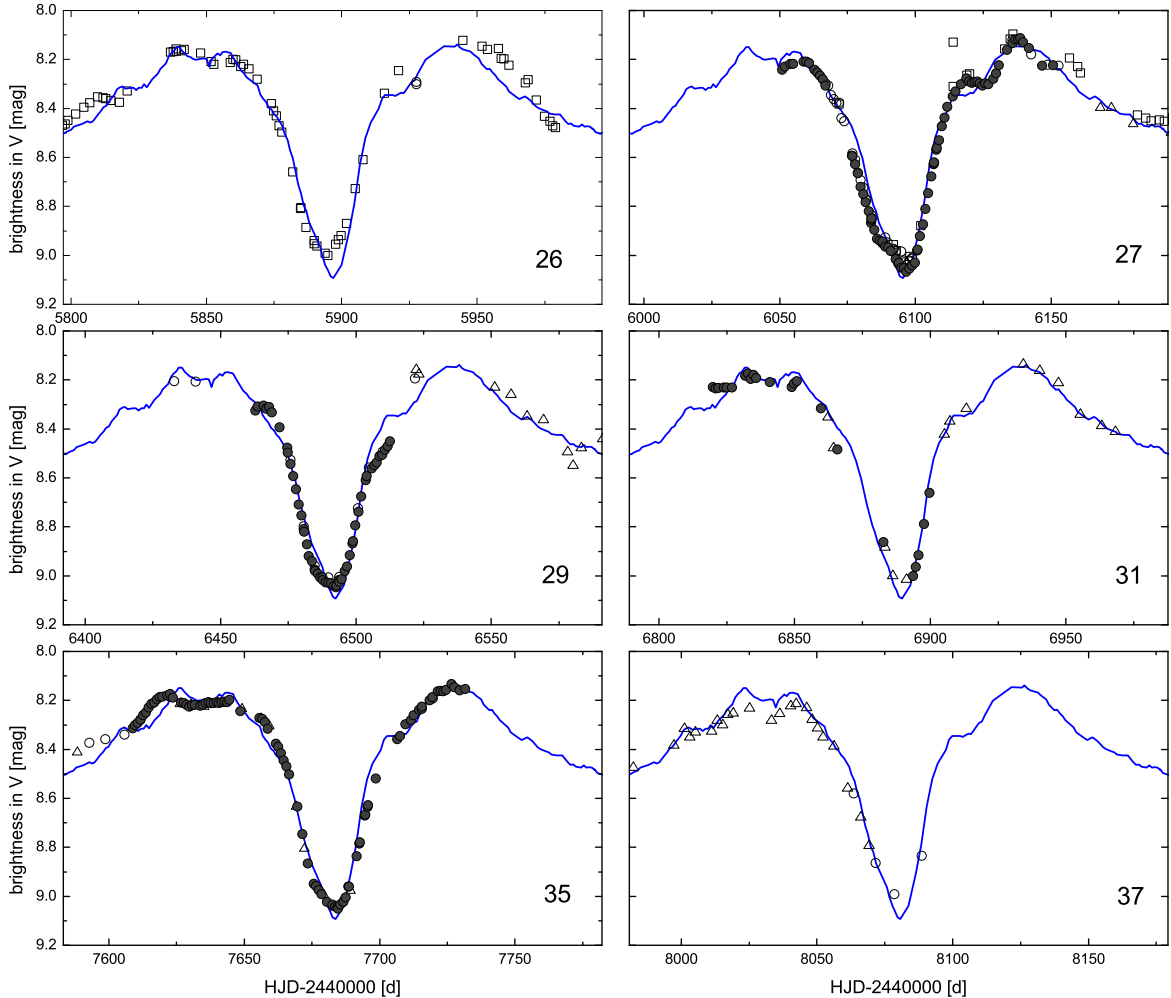,width=18.cm}
\caption{Time series of $V$ band photoelectric measurements of W Crucis,
  all available data sets are shown; filled circles stand for Geneva
  measurements (this work), open squares for Marino et al.\ 
(1988),  
  open triangles for Pazzi (1993), and open
  circles for ESO (Manfroid et al.\ 1991; Sterken et al.\ 1993).
  Full line represent
  synthetic light curves from our solution which are derived from
  observations made in a single cycle no.\ 33 (aprox.\ HJD 2447100 --
  2447400). Cycles are designated according the zero epoch in Kohoutek's 
(1988)
 ephemeris.}
\end{figure*}

\setcounter{table}{1}

\begin{table} %
\caption{Absolute parameters for the components and an accretion
 disk in W Cru. Quantities are derived from solutions made from $V$ band
light curves.}
\begin{tabular}{lrrr}
\hline
Parameter              &   Zo{\l}a       &     This   \\
                       &    (1996)       &     work \\
\hline
Mass ratio, $q$                      & 0.16   &  0.19    \\
Inclination, $i$                     & 87.8   &  88.2    \\
Mass of gainer, $M_1$ [M$_\odot$]    & 7.82   &  8.2  \\
Mass of loser, $M_2$ [M$_\odot$]     & 1.25   &  1.6  \\
Radius of loser, $R_2$ [R$_\odot$    & 76     &   71  \\
Disc outer radius, $R_d$ [R$_\odot$] & 132    &  124 \\
Disc thickness, $z_d$ [R$_\odot$]    & 14.5   &  17  \\
Disc outer temp., $T_{d,out}$ [K]    & 1500   &  3600 \\
\hline
\end{tabular}

\end{table}

\section{Binary parameters}

As shortly described in Sect.~1, both Zo{\l}a (1996) and Daems (1998)
have analyzed the light curves of W Cru with a disk model. Zo{\l}a
used mean light curves constructed from Marino et al.\ (1988), and
Pazzi (1993) photoelectric photometry, respectively. Daems study was
based on then unpublished Geneva photometry. Both studies have arrived
to a quite consistent set of the parameters. While in our initial
calculations we have started with the parameters which have cover very
broad range in the parameter space, it was immediately clear that
solutions will be in the narrow range around the values specified by
Zo{\l}a, and Daems, respectively.

Since we are looking for the optimal set of the parameters in a
multi-dimensional parameter space it is recommended to fix as many
parameters as possible. Therefore, we assigned theoretical values to
gravity brightening, limb darkening, and albedo coefficients.  The
effective temperature of the G supergiant, only visible stellar
component, has been derived by Daems (1998) from the spectral energy
distribution in the broad wavelength span, from IUE UV to optical and
IR photometry obtained at ESO. He has arrived to $T_{2,\rm eff} =
5500$ K.
   
Since W Cru is a single-lined SB and only the mass-function is known,
mass ratio has to be derived from the light curves. In our, as in
previous studies, a conservative assumption that mass-loosing
component is filling its critical Roche lobe, is assumed.  Therefore,
its size is determined by mass ratio. This fact makes determination of
the mass-ratio more easy through ellipticity effect. Otherwise, mass
ratio will be poorly determined.

Whatever initial set of parameters used, the solution has converged to
the case with no primary visible at all. The, final search for the
optimal set of the parameters was made with the remaining two binary
parameters: the mass ratio $q$, inclination of the orbit $i$, and the
following three disk parameters; disk outer temperature $T_{d, out}$,
angle $\beta_d$ defining disk semithickness, and disk outer radius
$r_{d,out}$.

The optimal solution, as defined by $\chi^2$ minimization with the
Genetic Algorithm (see Sect. 4.3 for details) has given for the mass
ratio $q = 0.19$, and inclination $i = 88\fdg2$. This should be
compared to $q = 0.16$, and $i = 87\fdg8$ derived by Zo{\l}a, and $q =
(0.175; 0.160)$, and $i = (88\fdg2; 87\fdg6)$ derived by Daems
depending on cycle.

In combination with the mass-function $f(m) = 5.83$ R$_\odot$ (Woolf
1962) for given mass ratio and inclination one can calculate the
masses of the components. We get, $M_1 = 8.2$ M$_\odot$, and $M_2 =
1.6$ M$_\odot$.  The sum of the masses, with the period which is
known, gives the separation of the components $A = 306$ R$_\odot$.
Since in the semi-detached configuration a relative size of the
Roche-lobe filling component is uniquely determined by the system's
mass ratio the radius of the supergiant in W Cru is $R_2 = 80$
R$_\odot$.

The optimal geometrical disk parameters are: $r_d = 0.405$, and
$\beta_d = 7\fdg8$ which translate to $R_d = 124$ R$_\odot$, and the
semithickness at the outer disk edge $z_{d, out} = 17$ R$_\odot$.  The
later parameter also defines an upper limit on the radius of the
invisible component $R_1 < 17$ R$_\odot$. This supports the hypothesis
that the primary component which is completely hidden by a thick
accretion disk, is a mid-B MS star. Tabulation of Harmanec (1988) for
the B-type MS star of mass about 8 M$_\odot$ gives $R \sim 4$
R$_\odot$, hidden well inside thick disk we have found, in particular
for almost edge-on view. Fig.~3 is showing time series plot of all
available photoelectric data along synthesized light curve on the
basis of our solution from Tab.~2. Intrinsic variations, as discussed
in Sect.~3, are clearly seen as due to variations in disk properties,
both radiative and geometrical.

We have attempted to compute solution for the $U$-band light curve.
While keeping binary and stellar parameters at the same values as in
the $V$-band solution, only disk parameters were allowed to be
adjusted. As expected, disk parameters differ for different passbands
with the disk more extended both in radial and vertical direction and
with a higher disk-rim temperature in the $U$-band ($R_d = 185$
R$_\odot$, $z_d = 22$ R$_\odot$, and $T_{\rm d, edge} = 4600$ K).
This is not surprising since both minima in $U$-band light curve are
almost the same depth and slightly more broader than in $V$-band light
curve (c.f.~Fig. 2).

\begin{figure}[!th]
\psfig{file=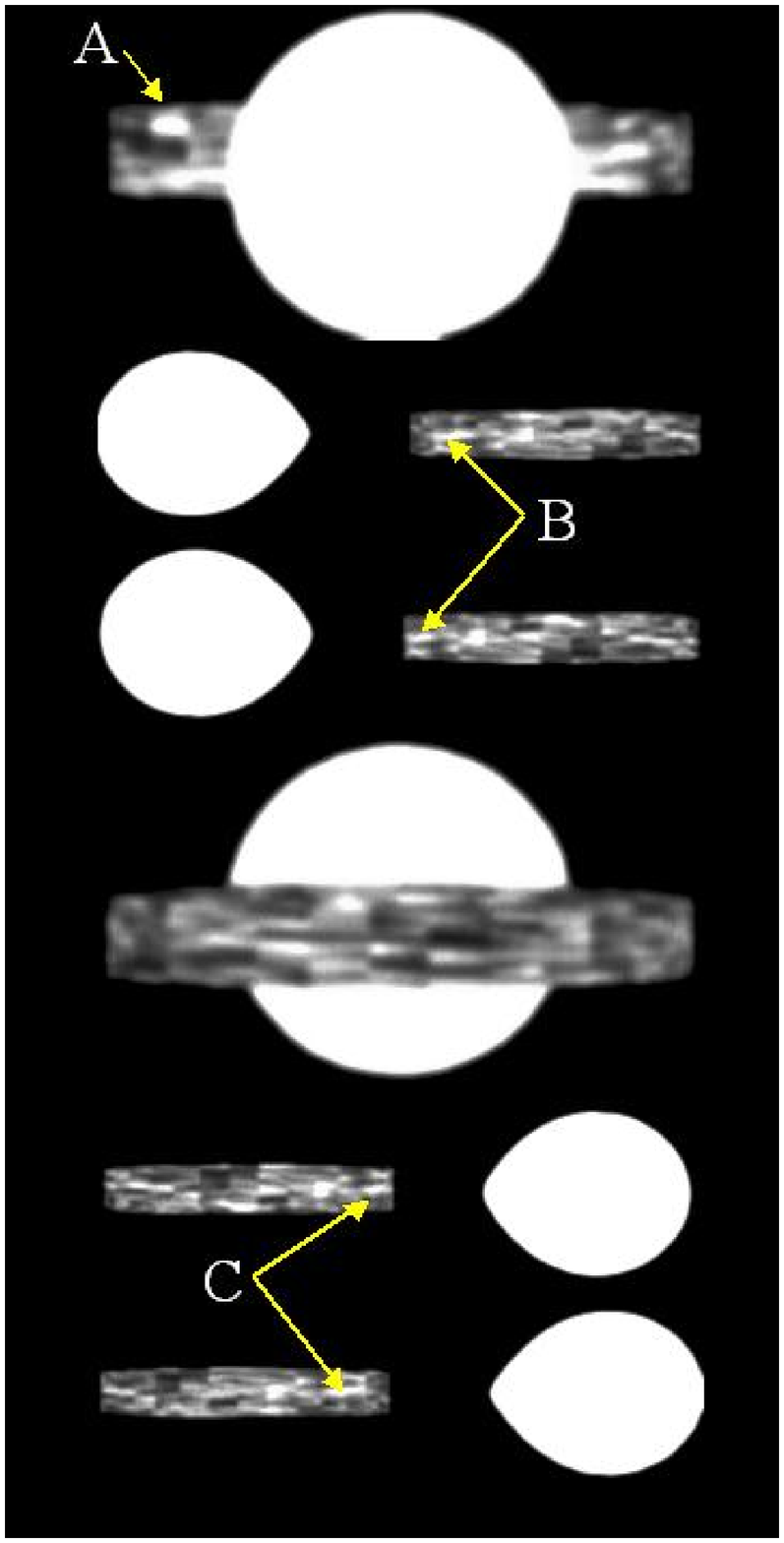,width=8.8cm}   
\caption{Reconstructed disk rim images shown in the phases 
 0.0, 0.25, 0.30, 0.50, 0.75, and 0.80 (from top to bottom)
in $V$ photometric passband. Corresponding synthetic light curves
 for the system with this accretion disk image is shown in upper
pannel of the Fig.\ 2. Some features are identified which produces 
bumps in the light curves at the phases 0.2, 0.3, and 0.65, 
respectively. }
\end{figure}

\begin{figure}[!h]
\psfig{file=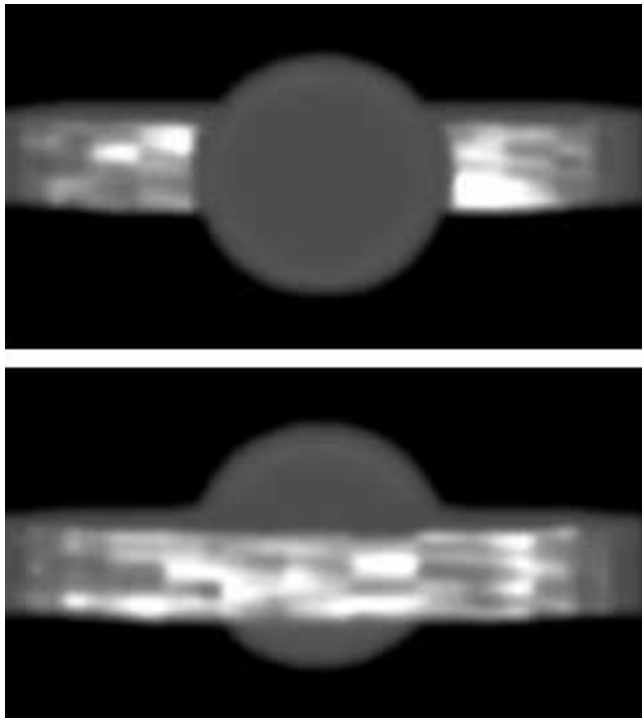,width=8.5cm} 
\caption{Reconstructed disk rim images shown in the phases of 
the 
minima 
in $U$ photometric passband. Corresponding synthetic light curves
 for the system with this accretion disk image is shown in bottom
pannel of the Fig.\ 2. Radial extension of the disk is larger
for  $U$ passband than $V$ passband. Some similarities of disk
rim images in both passbands are seen for the primary (deeper)
minimum, phase 0.50 (upper pannel), but it is not so clear for
the secondary minimum, phase 0.0, which is considerably deeper
inthe $U$ passband light curve comparing to the depth in $V$
passband (c.f.~Fig.\ 2). }
\end{figure}

Zo{\l}a (1996) calculated a disk edge temperature using Smak's (1992)
formula. Depending on the light curves used for the photometric
solution he obtained $T_{\rm d, edge} = 1500$ K, and 2000 K,
respectively.  Disk rim temperature derived in our modeling differs
from Zo{\l}a's estimates, and in particular for the U-band is
considerably larger.  Even seen edge-on the disk is an important
source of radiation; as is seen in Fig. 2 the minima are almost equal.
It is an eclipsing body producing a deep primary eclipse, and quite a
long secondary eclipse without visible eclipse-shoulders.

\section{An image of disk}

An optimal set of the binary and disk parameters give only the first
approximation to the light curve. Symmetric model of the binary can by
no means give a fit to the light curves which exhibit a complex
structure with asymmetric ascending and descending eclipse branches,
unequal maxima, and a number of bumps/humps in the light curve
(Fig.~3).  This was a serious problems faced by previous modelers as
already described in Sect.~1. Once an optimal set of the parameters is
found we can proceed with the second step in minimization (Sect. 4.3).
In this step our minimization procedure has to recover the surface
brightness distribution of an accretion disk. The disk is divided in
a grid of triangles. In looking for an optimal map of the disk
brightness distribution the intensities of triangles could be varied.
The optimal map is a reconstructed disk image. The information
contained in the complete light curve is used, not only the parts
covered in the eclipse (in the case of W Cru this will be the
secondary eclipse).

In Fig.~4 reconstructed disk images are shown for several important
phases in $V$ spectral passband.  Since
inclination is high we see almost a disk rim. The disk rim appears
rather clumpy with nonuniform brightness distribution. It is difficult
to speak on some structures but on the image in phase 0.0 when disk
rim is visible in full extension some kind of long 'filamentary' or
'stream' structures might be recognized. There is an impression that
they are slightly tilted to the orbital plane. Similar clumpy
structures are revealed from $U$-band light curve modeling (Fig.~5).
Initial disk parameters are listed in Tab.~1, and an image
reconstruction is obtained in the same procedure as was elaborated for
the $V$-band.

Hydrodynamical calculations of the gas flow in the close binaries have
been performed focusing mostly on the study of accretion disks in
cataclysmic variables. However, Bisikalo et al.\ (2000) have made
calculations for the model which has resembled the long-period
interacting binary $\beta$ Lyrae. So far, this is the only
hydrodynamical simulations conducted for long-period interacting
binaries. Some important conclusions from their work are as follows:
1) the matter of the gas stream from the Roche-lobe filling component
is redistributed into two parts, one that forms a quasi-elliptic
accretion disk, and the other which forms a circumstellar envelope
around the mass-gaining component, 2) the gas stream approaches the
disk tangentially and does not lead to the formation of the usually
postulated `hot spot', 3) in fact, the interaction between the gas
from the circumstellar envelope and the stream results in the
formation of an intensive shock wave, and faces the accreting star,
and 4) a part of the gas from the circumstellar envelope is deflected
away from the orbital plane after interaction with the incoming stream
and leaves the system while the other one moves in the orbital plane,
encircles the accretor and then collides with the stream. It is
evident that hydrodynamical simulations of mass flow in the
interacting binaries like $\beta$ Lyrae have shown a variety of the
structures like accretion disk, circumstellar envelope, 'hot line' and
cool stream, halo.

Very recent polarimetric study of W Serpentis, a prototype of this
isolated class of the strongly interacting binary stars, has been
accomplished by Piirola et al.\ (2005). This system resembles many
similarities with W Cru described in the present work (an asymmetric
eclipse, large perturbations in the light curve also outside the
eclipse, color variations in and outside the eclipse, etc.). Their
detail modeling of the polarization measurements has shown complex
structure of the circumstellar envelope in which the primary (hot)
component is embedded. While photometric and spectroscopic studies
have revealed the existence of a geometrically and optically thick
accretion disk around the hot component, no disk is seen in polarized
light probably because multiple scattering and absorption effects in
the optically thick medium reduce the polarization of the light. The
authors have localized scattering 'spot' which is probably associated
with the optically thin polar regions. Similar effect in polarization
behavior would be produced by a jet emerging perpendicular to the disk
plane. It is obvious that Piirola et al.\ (2005) polarimetric study of
W Ser emerged to the model similar to $\beta$ Lyr as proposed by
Harmanec et al.\ (1996) from an interferometric study, and supported
by spectropolarimetric investigation by Hoffman et al.\ (1998).

Our disk image reconstruction is based on a model that includes only
an accretion disk and probably suffers from the absence of additional
structures, in particular stream of the matter from cool companion in
direction of more massive component and/or its, already, formed
accretion disk.  But pseudophotosphere of this optically thick disk
which has completely hidden hot component of the binary system has
undertaken its role in the emission of radiation.

\section{Conclusion}

Systematic photoelectric photometry of the long-period eclipsing
binary W Cru in 7-color Geneva system has been carried out at the
Swiss Telescope at La Silla from 1984-1989 and is, for the first time,
presented in this paper. Several cycles of this almost 200 days long
period binary system have been covered. Signatures of the still
present activity in this system are obvious: maxima of the light
curves are unequal, asymmetry is present in the eclipse branches,
humps/bumps are present in the light curve and migrate from
cycle-to-cycle, the width of the eclipses are varying. Previous
research on this binary is corroborated: modeling of the light curve
with an accretion disk model has revealed geometrically and optically
thick accretion disk in which a more massive and presumably hotter
component of mid-B type is completely hidden.

We have used 3D eclipse-mapping technique to reconstruct an accretion
disk image. Since the binary is viewed almost edge-on our map revealed an
image of accretion disk rim only. Its nonuniform brightness distribution
reveals  rather  clumpy structure. We shortly discussed reconstructed
disk image with hydrodynamical simulations of the matter flow in 
semi detached binaries which have shown that a number of features and
structures might be formed in such processes.

\begin{acknowledgements}

We acknowledge constructive comments from our referee Dr.~Petr Harmanec 
which improved this paper.
The observational part of this work has been partly supported by the
Swiss National Science Foundation. 
KP acknowledges financial support
from Croatian MZOS through research grant
\#0119254.

\end{acknowledgements}

\end{document}